\documentclass[journal]{new-aiaa}
\usepackage[utf8]{inputenc}

\usepackage{graphicx}
\usepackage{amsmath}
\usepackage[version=4]{mhchem}
\usepackage{siunitx}
\usepackage{subcaption}
\usepackage{longtable,tabularx}
\usepackage{xcolor}
\title{Coverage Area Determination for Conical Fields of View Considering an Oblate Earth}

\author{Marco Nugnes\footnote{Ph.D. Student, Department of Aerospace Science and Technology; marco.nugnes@polimi.it.} and Camilla Colombo\footnote{Associate Professor of Orbital Mechanics, Department of Aerospace Science and Technology; camilla.colombo@polimi.it.}}
\affil{Politecnico di Milano, 20156 Milan, Italy} 
\author{\vspace*{-0.4 cm} and \\ Massimo Tipaldi\footnote{R\&D Proposal Manager, Department of Engineering; mtipaldi@unisannio.it.}}
\affil{University of Sannio, 82100 Benevento, Italy}
\author{\vspace*{-0.2 cm} \small{DOI: 10.2514/1.G004156}}
\begin{document}

\maketitle
\vspace*{-0.7cm}
\begin{abstract}
This paper introduces a new analytical method for the determination of the coverage area modeling the Earth as an oblate ellipsoid of rotation. Starting from the knowledge of the satellite’s position vector and the direction of the navigation antenna line of sight, the surface generated by the intersection of the oblate ellipsoid and the assumed conical field of view is decomposed in many ellipses, obtained by cutting the Earth’s surface with every plane containing the navigation antenna line of sight. The geometrical parameters of each ellipse can be derived analytically together with the points of intersection of the conical field of view with the ellipse itself by assuming a proper value of the half-aperture angle or the minimum elevation angle from which the satellite can be considered visible from the Earth’s surface. The method can be applied for different types of pointing (geocentric, geodetic and generic) according to the mission requirements. Finally, numerical simulations compare the classical spherical approach with the new ellipsoidal method in the determination of the coverage area, and also show the dependence of the coverage errors on some relevant orbital parameters.
\end{abstract}

\section*{Nomenclature}

{\renewcommand\arraystretch{1.0}
\noindent\begin{longtable*}{@{}l @{\quad=\quad} l@{}}
\textbf{A}\textsubscript{rot} & rotation matrix from the geocentric equatorial frame to the local ellipse reference frame \\
\textbf{A}\textsubscript{321} & rotation matrix from the geocentric equatorial frame to the reference frame aligned with the line of sight\\
$a$  & ellipsoid semi-major axis, km\\
$\tilde{a}$ & semi-major axis of the ellipse originated from the intersection of a generic plane and the oblate ellipsoid, km \\ 
$b$ &  ellipsoid semi-minor axis, km \\
$\tilde{b}$ & semi-minor axis of the ellipse originated from the intersection of a generic plane and the oblate ellipsoid, km \\ 
$d$ & distance of the intersection plane from the origin, km \\
$E$ & oblate Earth's eccentricity \\
$\mathbf{\hat{{e}}}$ & ellipse apse line direction \\
$f$ & Earth flattening factor \\
$H$ & satellite's height, km \\
$i$ & orbit inclination, rad \\
$m$  & angular coefficient \\
$\mathbf{\hat{n}}$ & unit vector normal to a generic plane \\
$\mathbf{\hat{o}}$ & line-of-sight opposite direction \\
$q$ & vertical intercept, km\\
$\textbf{R}$ & rotation matrix\\
{\emph{R}}\textsubscript{eq} & Earth equatorial radius, km \\
{\emph{R}}\textsubscript{pol} & Earth polar radius, km\\
$R_{\oplus}$ & mean Earth equatorial radius, km\\
$\mathbf{r}_{P}$ & position vector of a generic point on the local ellipse, km \\
$\mathbf{r}_{s/c}$ & spacecraft position vector, km \\
$\mathbf{r}_{T}$ & position vector of the local ellipse point of tangency with the conical field of view, km \\
$\mathbf{\hat{s}}$ & position vector of the ellipse's center originated from the intersection of a generic plane and the oblate ellipsoid, km \\
$\mathbf{\hat{u}}$ & semi-minor axis direction \\
$\alpha$ & angle between of a generic line and the apse line direction, rad\\ 
$\beta$ & angle between the line of sight and the apse line direction, rad \\
$\gamma$& intermediate angle, rad \\
$\Delta$ & discriminant of a second-order equation \\
$\varepsilon$ & elevation angle, rad \\
$\eta$  & half-aperture angle, rad \\
$\theta$ & local sideral time, s \\
$\Lambda$ & ground range angle, rad \\
$\lambda$   & geographic longitude, rad\\
$\nu$ & argument of latitude, rad \\
$\rho$ & slant range, km \\
$\phi$ & geographic latitude, rad \\
$\psi$ & roll angle, rad \\
$\Omega$ & right ascension of the ascending node, rad  \\
\newpage
\multicolumn{2}{@{}l}{\emph{Subscripts}} \vspace*{0.2 cm} \\
eq & equatorial\\
FOV & field of view \\
gc & geocentric \\
gd & geodetic \\
hor & horizon \\
int & intersection \\
pol & polar \\
$s/c$ & spacecraft \\
\end{longtable*}}

\section{Introduction}
\lettrine{T}{he} determination of the coverage area associated to a navigation antenna or an optical sensor having a conical field of view is important for the preliminary design of any space mission. This problem has been already analyzed in the classic literature such as in Vallado \cite{Vallado} and in Wertz and Larson \cite{SMAD}. Both Vallado and Wertz and Larson derive the formulations to compute the coverage area under the assumption of a perfect spherical Earth. Vallado takes as input the half-aperture angle of the instrument, assuming a conical field of view, for its proof. On the other hand, Wertz and Larson consider as starting point the elevation angle of the desired location on the Earth. The two approaches achieve the same final result.

The ellipsoidal representation of the Earth has been already analyzed for the missile trajectory computation: knowing the launch site and target point of a missile the path length, called ground range, is derived above the surface of the Earth. This problem was explored by Vincenty \cite{Vincenty}, who offered an iterative algorithm that is nowadays used for the ground range computation. Also Nguyen and Dixson \cite{Nguyen} and Escobal \cite{Escobal_Missile} solved this problem in an analytic way defining the elliptic curve over the surface of an oblate Earth model by the intersection of a plane that passes through the launch site, the target point, and the center of the oblate spheroid. Clearly in this case, the center of the intersected ellipse is also the center of the oblate spheroid, leading to a particular case and not a general solution of the problem. A complete analytical solution has been discussed by Maturi et al. \cite{Maturi}, where the center of the ellipse can be different from the center of the Earth. These examples take into account only the ground range computation starting from the launch site and the target point. 

As regards the computation of the coverage area associated to a satellite, Escobal \cite{Escobal} derives in his book a closed-form solution to the satellite visibility problem. The closed-form solution is a single transcendental equation in the eccentric anomalies corresponding to the rise and set times for a given orbital pass of a satellite under the assumptions of Keplerian motion and knowledge of satellite orbital elements, station coordinates and minimum elevation angle. The elevation angle is the angle between a satellite and the observer’s (ground station’s) horizon plane \cite{Cakaj}. Escobal \cite{Escobal} solves also the transcendental equation by using numerical methods once per satellite period, which is faster than determining the value of the elevation angle of the satellite with respect to a ground station for each time instant. 

Lawton \cite{Lawton} developed a method to solve for satellite-satellite and satellite-ground station visibility periods considering an oblate Earth defining a new visibility function based on the vertical distance above the plane tangent to a ground station by using a Fourier series. Exploiting the sinusoidal nature of the visibility curve generated by satellites with orbital eccentricities less
than 0.1, he determines the local periodicity of this curve and then uses a numerical search to locate rise and set times. This method works well for low eccentricity orbits, but fails for more elliptical orbits because the visibility waveform becomes aperiodic.
Alfano et al. \cite{Alfano} extend the use of the visibility function defined by Lawton \cite{Lawton} for all the types of orbit, presenting an algorithm which exploits a parabolic blending, a space curve modeling, to construct the waveform of the visibility function. 
The determination of the rise and set times is the starting point for the coverage analysis since they enclose the region in which the satellite is visible.

In his report, Walker \cite{Walker} derives circular orbital patterns providing continuous whole Earth coverage modeling the Earth as a perfect sphere. This assumption can be considered reasonable if the satellites' altitude is low since the spherical geometry and the ellipsoidal one are not so different.
Ma and Hsu \cite{Ma} proposed a solution for the exact design of a partial coverage satellite constellation over an oblate Earth, that is, the coverage of certain regions of the Earth with gap times in coverage no longer than some specified maximum time, which is based on a visibility function defined in Chylla and Eagle \cite{Chylla}, analogous to the one introduced by Escobal \cite{Escobal}. 
This method cannot be efficient for a Global Navigation Satellite System (GNSS) where the coverage is supposed to be global, and there is a significant number of satellites. The problem in using the visibility function is the resolution of the transcendental equation to determine the region visible from each satellite. Indeed, to the best of our knowledge, no analytical formulations relating the elevation angle or the half-aperture angle, modeling the navigation signal as a conical field of view, exist.

In this paper a new analytical method for the determination of the coverage area, having as input the line of sight of the satellite, the half-aperture angle of the conical field of view, and the satellite position vector considering the Earth as an oblate ellipsoid of rotation, is presented. Unlike the previous papers where the rise and set times of a satellite are evaluated from a generic ground station, this approach starts from the position of a generic spacecraft on its orbit and derives all the locations on the Earth's surface within its field of view. Moreover, it is also numerically shown that, while for the spherical case the locus of points on the Earth's surface is a circle, for the oblate ellipsoid case there is no planar geometric line containing all the points.  

This paper is organized as follows: the intersection of a generic plane with an oblate ellipsoid of rotation is summarized in Sec. II. In Sec. III the derivation of the analytic formulation to compute the coverage area is presented, and in Sec. IV the method is applied for different orbital pointing scenarios. Numerical simulations to validate the models and the results are carried out in Sec. V together with a direct comparison between the spherical approach and the new ellipsoidal approach. Finally, Sec. VI concludes the paper.

\section{Background}

In this section the state-of-the-art for the determination of the coverage area is described starting from the classical derivation considering the Earth as a perfect sphere (see Appendix) and moving to the intersection of a generic plane with the Earth modeled as an oblate ellipsoid of rotation.
Apart from the derivation of the intersection of the oblate ellipsoid of rotation with a generic plane that is taken from \cite{Maturi}, the computation of the intersection points of a conical field of view with the Earth's surface is an original work of this paper (see Sec. III). The main difference with respect to the spherical approach is that the Earth is assumed as an oblate ellipsoid of rotation. This can be considered a more refined model with respect to the previous one since the oblate ellipsoid approximates better the shape of the Earth.

\begin{figure}[!ht]
\centering
\includegraphics[scale = 0.8]{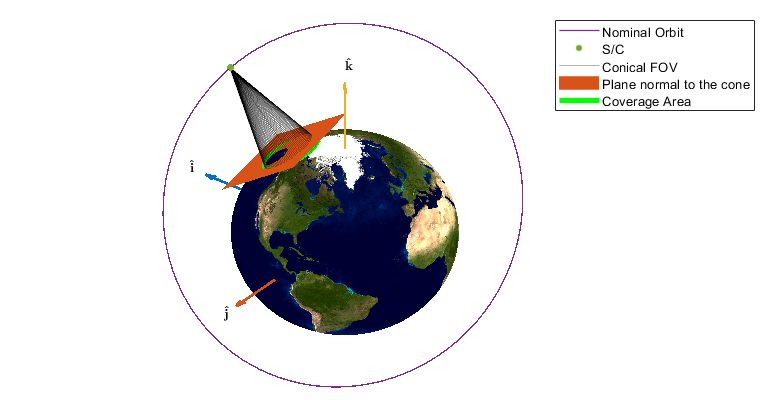}
\caption{Projection of a conical field of view on an oblate ellipsoid.}
\label{Fig_1}
\end{figure}

Also in this case, the navigation signal is assumed to be  extending as a cone \cite{Vallado} from the center of mass of the spacecraft with a given half-aperture angle $\eta$. Whereas in the spherical approach the intersection of a plane with the Earth is a circle, this case is different as shown in Fig. \ref{Fig_1}, because the coverage area is a three-dimensional surface and is not contained in a plane normal to the conical field of view.

The first step is to prove that the trace of the oblate ellipsoid on a generic plane is an ellipse \cite{Maturi}, because in this way it is possible to move from a 3D problem to a planar one as in the spherical Earth.
The general equation of a plane is given by:
\begin{equation}
n_1x + n_2y + n_3z = d
\label{Eq_1}
\end{equation} 
where $\mathbf{\hat{n}} = [n_1,n_2,n_3]$ is a unit vector normal to the plane and $d$ is the distance of the plane from the origin of the coordinate frame. In this case the vector normal to the plane is coincident with the line of sight of the navigation signal, while the origin of the coordinate frame is the Earth's center.
The equation of the oblate spheroid can be written as:
\begin{equation}
\frac{x^2 + y^2}{a^2} + \frac{z^2}{b^2} = 1
\label{Eq_2}
\end{equation} 
where $a$ is the semi-major axis and $b$ is the semi-minor axis of the oblate ellipsoid. In the specific case the semi-major axis is the Earth equatorial radius and the semi-minor axis is the Earth polar radius.
The trace of the oblate ellipsoid in the generic plane is computed deriving the expression of $z$ in Eq.\hspace{1.5 mm}(\ref{Eq_1}) and then substituting in Eq.\hspace{1.5 mm}(\ref{Eq_2}). For $n_3 \neq 0$, it is possible to write:
\begin{equation}
z = \frac{d - n_1x - n_2y}{n_3}
\label{Eq_3}
\end{equation}  

If $n_3 = 0$, then $\mathbf{\hat{n}}$ is parallel to the equatorial plane and perpendicular to the polar axis. Substituting Eq.\hspace{1.5 mm}(\ref{Eq_3}) in Eq.\hspace{1.5 mm}(\ref{Eq_2}):
\begin{equation}
\frac{x^2 + y^2}{a^2} + \frac{1}{b^2n^2_3}\left(d^2 + n^2_1x^2 + n^2_2y^2 - 2dn_2x - 2dn_2y + 2n_1n_2xy\right) = 1
\label{Eq_4}
\end{equation} 

Developing the products and rearranging the equation, the result is the following:
\begin{equation}
\scalebox{1}{$(a^2n^2_1 + b^2n^2_3)x^2 + (a^2n^2_2 + b^2n^2_3)y^2 + 2a^2n_1n_2xy - 2dn_2a^2x - 2dn_2a^2y - a^2b^2n^2_3 + a^2d^2 = 0$}
\label{Eq_5}
\end{equation} 
which is the equation of a conic. The general equation of a conic is:
\begin{equation}
Ax^2 + By^2 + 2Gx + 2Fy + 2Hxy + C = 0
\label{Eq_6}
\end{equation} 

This equation represents an ellipse or a circle if its discriminant, $\Delta$, is less than 0. The discriminant of Eq.\hspace{1.5 mm}(\ref{Eq_5}) is:
\begin{equation}
\Delta = {(a^2n_1n_2)}^2 - (a^2n^2_1 + b^2n^2_3)(a^2n^2_2 + b^2n^2_3)
\label{Eq_7}
\end{equation} 

After some manipulations, the result is obtained as:
\begin{equation}
\Delta = -b^2n^2_3(a^2n^2_1 + a^2n^2_2 + b^2n^2_3) < 0
\label{Eq_8}
\end{equation} 
which is always less than zero. This means that the intersection of the oblate ellipsoid with a generic plane is an ellipse or a circle. 

After showing that the intersection of an oblate ellipsoid on a generic plane is represented by an ellipse, it is necessary to compute the value of the geometric quantities of this ellipse following \cite{Maturi}: the center of the ellipse, its semi-major axis and its semi-minor axis.
Without going into details, the previous quantities are analytically computed with the following formulas taken from \cite{Maturi}. The center of the local ellipse with respect to the origin of the reference frame is computed with Eq. (\ref{Eq_9}).
\begin{equation}
\mathbf{s} = \left[\frac{n_1d}{1 - E^2n^2_3}, \frac{n_2d}{1 - E^2n^2_3}, \frac{(b^2/a^2)n_3d}{1 - E^2n^2_3}\right]
\label{Eq_9}
\end{equation} 

\vspace{0.2 cm}
\noindent where $E$ is the oblate Earth's eccentricity. Naturally, if the generic plane passes through the center of the Earth, then: 
\begin{equation*}
d = 0
\end{equation*}
and, consequently, $\mathbf{s} = [0,0,0]$; that is, the center of the oblate spheroid will also be the center of the ellipse. If the conical field of view of the navigation signal is divided into different planes having in common the satellite's line of sight, each plane will generate an ellipse hosting two limiting points for the coverage area, which is the output of the analysis.  
The semi-major axis of the local ellipse is defined by:
\begin{equation}
\tilde{a} = a\sqrt{1 - \frac{d^2}{a^2(1 - E^2n^2_3)}}
\label{Eq_10}
\end{equation}	
with $\tilde{a}$ the magnitude of the semi-major axis of the ellipse and $a$ the semi-major axis of the oblate ellipsoid of rotation (i.e., the Earth equatorial radius). The direction of the generic unit vector $\mathbf{\hat{e}}$ to be aligned with the apse line of the intersection ellipse is given by:
\begin{equation}
\mathbf{\hat{e}} = \frac{1}{\sqrt{n^2_1 + n^2_2}}\left[n_2,-n_1,0\right]
\label{Eq_11}
\end{equation}

The last quantity to be defined is the semi-minor axis of the intersection ellipse. The magnitude of the semi-minor axis is given by:	
\begin{equation}
\tilde{b} = b\frac{\sqrt{1 - d^2/a^2 - E^2n^2_3}}{1 - E^2n^2_3}
\label{Eq_12}
\end{equation}
with $\tilde{b}$ representing the semi-minor axis of the ellipse and $b$ the semi-minor axis of the oblate ellipsoid of rotation (i.e., Earth polar radius).
The direction of the generic unit vector $\mathbf{\hat{u}}$ to be aligned with the semi-minor axis is given by:
\begin{equation}
\mathbf{\hat{u}} = \frac{1}{\sqrt{n^2_1 + n^2_2}}\left[n_1n_3, n_2n_3, -(n^2_1 + n^2_2)\right]
\label{Eq_13}
\end{equation}	

\begin{figure}[ht!]
\centering
\includegraphics[scale = 0.7]{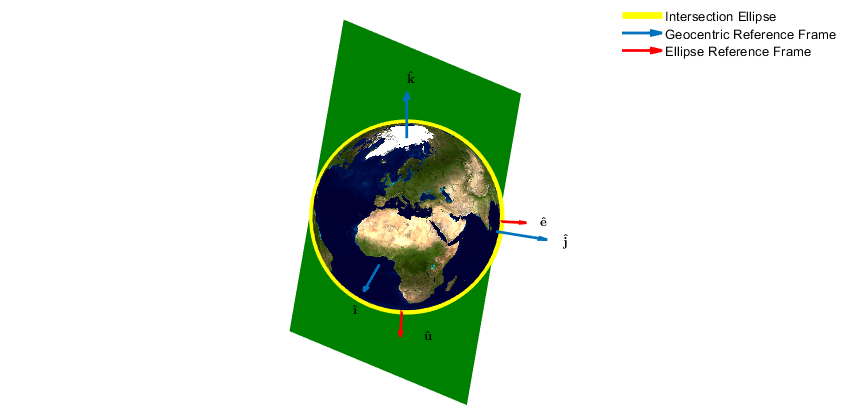}
\caption{Representation of the intersection ellipse with the oblate ellipsoid of rotation.}
\label{Fig_2}
\end{figure}

Figure \ref{Fig_2} summarizes the geometric parameters defined in this section putting in evidence the intersection ellipse with the oblate ellipsoid of rotation. 

\section{Computation of the Coverage Parameters for an Oblate Earth}

\begin{figure}[ht!]
\centering
\includegraphics[scale = 0.8]{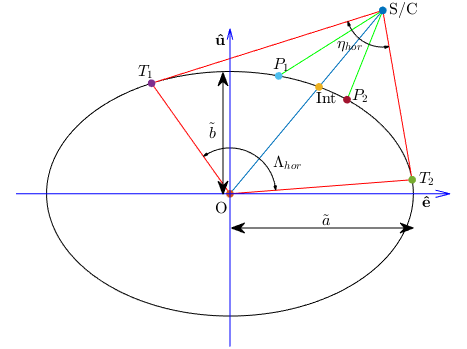}
\caption{Representation of the elliptical field-of-view problem.}
\label{Fig_3}
\end{figure}

In the previous section all the geometric quantities of the intersection ellipse have been determined. This gives the possibility to move from a 3D view to a planar view and determine the same quantities computed using the spherical approach. The new problem representation is identified in Fig.\hspace{1.5 mm}\ref{Fig_3}.
	
It is possible now to define a local reference frame in the plane of the ellipse obtained by the intersection of the oblate ellipsoid of rotation and a generic plane, where $\mathbf{\hat{e}}$ is the unit direction vector along the semi-major axis and $\mathbf{\hat{u}}$ is the unit direction vector along the semi-minor axis. The equation of the ellipse in this planar reference frame is:	
\begin{equation}
\frac{e^2}{\tilde{a}^2} + \frac{u^2}{\tilde{b}^2} = 1  
\label{Eq_14} 
\end{equation}	
where $e$ is just the first coordinate of a generic point in the local reference frame and it has not to be confused with the magnitude of the unit vector $\mathbf{\hat{e}}$. The first parameter to be determined is the horizon delimited by the points $T_1$ and $T_2$ given by the tangents to the ellipse drawn from the center of mass of the spacecraft. Let $m$ and $q$ be, respectively, the generic angular coefficient and the vertical intercept of a line in the local reference frame. The equation of a line in the local reference frame in the explicit form is:
\begin{equation}
u = me + q
\label{Eq_15}
\end{equation}

The two tangents in $T_1$ and $T_2$, as shown in Fig. \ref{Fig_3}, have in common the satellite's center of mass whose coordinates are known if the satellite's position vector in the inertial frame is known.
\begin{equation}
\mathbf{r}_{s/c} = [e_{s/c}, u_{s/c}, 0] 
\label{Eq_16}
\end{equation}

The two points of tangency are obtained as intersection of the equation of the tangent and the equation of the local ellipse.	
\begin{equation}
\begin{cases}
u - u_{s/c} = m(e - e_{s/c})\\ 
\frac{\scalebox{1}{\hspace{-0.15 cm}$e^2$}}{\scalebox{1}{$\tilde{a}^2$}} + \frac{\scalebox{1}{$u^2$}}{\scalebox{1}{$\tilde{b}^2$}} = 1   
\end{cases}
\label{Eq_17}
\end{equation}

Developing the equations, it is possible to get to the following system:	
\begin{equation}
\begin{cases}
u = me - me_{s/c} + u_{s/c}\\ 
\scalebox{1}{$({\tilde{b}^2} + \tilde{a}^2m^2)e^2 - 2\tilde{a}^2m(me_{s/c}-u_{s/c})e + (\tilde{a}^2m^2e^2_{s/c} + \tilde{a}^2u^2_{s/c} - 2\tilde{a}^2me_{s/c}u_{s/c} - \tilde{a}^2\tilde{b}^2) = 0 $ }
\end{cases}
\label{Eq_18}
\end{equation}	

The discriminant, $\Delta$, associated to the second of Eq. (\ref{Eq_18}) is the following:	
\begin{equation}
\frac{\Delta}{4} = (\tilde{a}^4\tilde{b}^2 - \tilde{a}^2\tilde{b}^2e^2_{s/c})m^2 + 2\tilde{a}^2\tilde{b}^2e_{s/c}u_{s/c}m + (\tilde{a}^2\tilde{b}^4 - \tilde{a}^2\tilde{b}^2u^2_{s/c})
\label{Eq_19}
\end{equation}	

From the value of the discriminant it is possible to deduce the type of solution. To have the tangency, the value of $\Delta$ needs to be equal to zero. This leads to a second-order equation in the unknown $m$.	
\begin{equation}
(\tilde{a}^2 - e^2_{s/c})m^2 + 2e_{s/c}u_{s/c}m + (\tilde{b}^2 - u^2_{s/c}) = 0
\label{Eq_20}
\end{equation} 

The roots of Eq.\hspace{1.5 mm}(\ref{Eq_20}) are given by these two expressions:	
\begin{equation}
m_{T_1} = \frac{-e_{s/c}u_{s/c} + \sqrt{\tilde{a}^2u^2_{s/c} + \tilde{b}^2e^2_{s/c}-\tilde{a}^2\tilde{b}^2}}{(\tilde{a}^2 - e^2_{s/c})} \hspace{1 cm}
m_{T_2} = \frac{-e_{s/c}u_{s/c} - \sqrt{\tilde{a}^2u^2_{s/c} + \tilde{b}^2e^2_{s/c}-\tilde{a}^2\tilde{b}^2}}{(\tilde{a}^2 - e^2_{s/c})}
\label{Eq_21}
\end{equation}

Once the value of the angular coefficient of the two tangents is known, it is possible to compute the points of intersection with the ellipse, whose coordinates are:	
\begin{equation}
e_{T_1} = \frac{\tilde{a}^2m_{T_1}(m_{T_1}e_{s/c} - u_{s/c})}{(\tilde{b}^2 + \tilde{a}^2m^2_{T_1})} \hspace{1 cm}
e_{T_2} = \frac{\tilde{a}^2m_{T_2}(m_{T_2}e_{s/c} - u_{s/c})}{(\tilde{b}^2 + \tilde{a}^2m^2_{T_2})}
\label{Eq_22}
\end{equation}

From the equation of the tangents it is possible to derive the other coordinates:	
\begin{equation}
u_{T_1} = m_{T_1}e_{T_1} - m_{T_1}e_{s/c} + u_{s/c} \hspace{1 cm}
u_{T_2} = m_{T_2}e_{T_2} - m_{T_2}e_{s/c} + u_{s/c}
\label{Eq_23}
\end{equation}	

In this way the vectors associated to the points of intersection of the tangents with the local ellipse are defined:	
\begin{equation}
\mathbf{r}_{T_1} = [e_{T_1},u_{T_1},0]    \hspace{1 cm}   \mathbf{r}_{T_2} = [e_{T_2},u_{T_2},0]
\label{Eq_24}
\end{equation}

The equivalent of the horizon-ground range angle ($\Lambda$\textsubscript{hor}) computed for the spherical case can be identified also for the ellipsoidal case considering the angle between these two vectors:	
\begin{equation}
\cos(\Lambda\textsubscript{hor}) = \frac{\mathbf{r}_{T_1}\cdot\mathbf{r}_{T_2}}{{r}_{T_1}{r}_{T_2}}
\label{Eq_25}
\end{equation}	

The horizon-boresight angle, $\eta$\textsubscript{hor}, can be determined in a similar manner knowing the position vector of the two points of tangency and the satellite's center of mass, because it is possible to compute the relative position of the spacecraft with respect to the two points of tangency. The angle between the two relative positions will represent the horizon-boresight angle. 
\begin{equation}
\cos(\eta\textsubscript{hor}) = \frac{(\mathbf{r}_{s/c} - \mathbf{r}_{T_1})\cdot\mathbf{r}_{s/c}}{\vert \mathbf{r}_{s/c} - \mathbf{r}_{T_1}\vert \vert\mathbf{r}_{s/c}\vert}
\label{Eq_26}
\end{equation}

\begin{figure}[ht!]
	\centering
	\includegraphics[scale = 0.8]{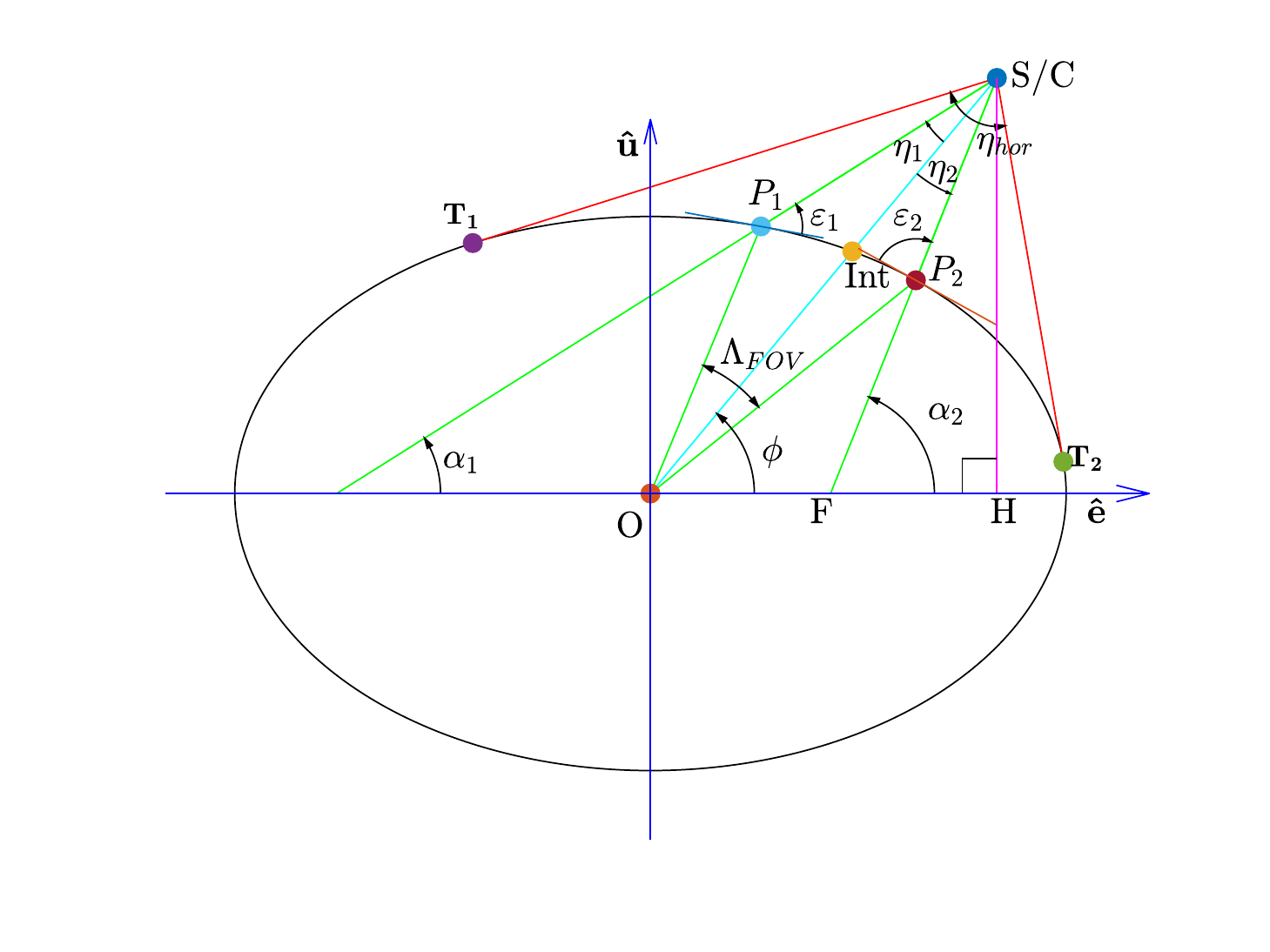}
	\caption{Highlight of the elliptical geometry.}
	\label{Fig_4}
\end{figure}

The next step is to derive the ground range angle that is associated to the aperture angle of the instrument on-board of the spacecraft. Using always the assumption of the conical field of view, three different situations may be identified: geocentric (line of sight coincident with the conjunction with the center of the Earth), geodetic (line of sight aligned with the local vertical to the Earth's surface), and generic (moving line of sight) pointing. The difference between the three methods from the geometrical point of view is the bisector line of the aperture angle. This subject will be better explored in the next section. For now, it is sufficient to assume that $\eta$ is the half-aperture angle with respect to a given line that forms an angle $\phi$ with the $\mathbf{\hat{e}}$ axis as visible from Fig.\hspace{1.5 mm}\ref{Fig_4}.
To determine the angular coefficient needed to compute the equations of the two secants in $P_1$ and $P_2$ intersecting the local ellipse, it is sufficient to draw the perpendicular to the $\mathbf{\hat{e}}$ axis from the satellite's center of mass. In this way, it is possible to make some geometrical considerations using the right triangles having as vertices $O$, $H$ and the satellite's center of mass and $F$, $H$ and satellite's center of mass. Let $\alpha_1$ and $\alpha_2$ be the angles formed by the two secants and the $\mathbf{\hat{e}}$ axis, respectively. It is possible to prove that: 
\begin{equation}
\label{Eq_27}
\begin{gathered}
	\alpha_1 = \phi - \eta \hspace{1 cm}
	\alpha_2 = \phi + \eta
\end{gathered}
\end{equation}

The angular coefficients of the two secants are simply the tangents of the two angles, visible in Fig.\hspace{1.5 mm}\ref{Fig_4}, defined before.
\begin{equation}
	m_{P_{1}} = \tan(\alpha_1)   \hspace{1 cm}    m_{P_{2}} = \tan(\alpha_2)
	\label{Eq_28}
\end{equation}

The next step is to write the generic equation for a line having as angular coefficient $m_{P_{1}}$ and $m_{P_{2}}$ and to solve a system with the equation of the local ellipse. The result is the same second-order equation (\ref{Eq_18}) with the difference that this time the discriminant will be or higher or lower than zero. 
In the latter case, no intersection with the ellipse are present.
Let the discriminant $\Delta$ be higher than 0, therefore there are two distinct solutions for each angular coefficient and so for each secant.
\begin{equation}
\label{Eq_29}
\begin{gathered}
e_{P_1} = \frac{\scalebox{1}{$\tilde{a}^2m_{P_1}(m_{P_1}e_{s/c}-u_{s/c}) \pm \sqrt{(\tilde{a}^4\tilde{b}^2 - \tilde{a}^2\tilde{b}^2u^2_{s/c}){m^2_{P_1}} + 2\tilde{a}^2\tilde{b}^2e_{s/c}u_{s/c}m_{P_1} + (\tilde{a}^2\tilde{b}^4 - \tilde{a}^2\tilde{b}^2u^2_{s/c})}$}}{\scalebox{1}{$\hspace{2 cm}(\tilde{b}^2 + \tilde{a}^2m^2_{P_1})$}} \\	
e_{P_2} = \frac{\scalebox{1}{$\tilde{a}^2m_{P_2}(m_{P_2}e_{s/c}-u_{s/c}) \pm \sqrt{(\tilde{a}^4\tilde{b}^2 - \tilde{a}^2\tilde{b}^2e^2_{s/c}){m^2_{P_2}} + 2\tilde{a}^2\tilde{b}^2e_{s/c}u_{s/c}m_{P_2} + (\tilde{a}^2\tilde{b}^4 - \tilde{a}^2\tilde{b}^2u^2_{s/c})}$}}{\scalebox{1}{$\hspace{2 cm}(\tilde{b}^2 + \tilde{a}^2m^2_{P_2})$}}
\end{gathered}
\end{equation}	

Of course, just two of these solutions are correct and these are the ones on the same side and closer to the center of mass of the spacecraft. To choose the right solution a distinction according to the quadrant has to be done and the values of $\alpha_1$ and $\alpha_2$ are involved. 
For example, in the first quadrant the value of $\alpha_1$ is always less than $\pi/2$. This means that once the right solution is identified, in the specific case the positive solution, it keeps the same sign. A different reasoning should be done for the other solution because the value of $\alpha_2$ starts lower than $\pi/2$ and then it becomes higher than $\pi/2$. The complete list of the cases together with the right solution to be picked is reported in Table \ref{Tab_1}.\\
\begin{centering}
\begin{table}[ht!]
\centering
\caption{Decision table to select the right solutions.}
\label{Tab_1}
\begin{tabular}{cccc}
\hline \hline 
\multicolumn{2}{c}{First Quadrant}                                                            & \multicolumn{2}{c}{Second Quadrant}                                                          \\ \hline
if $\alpha_2$ \textless{}= $\pi/2$  & if $\alpha_2$ \textgreater \,$\pi/2$  & if $\alpha_1$ \textless{}= $\pi/2$  & if $\alpha_1$ \textgreater\, $\pi/2$ \\
$e_{P_1}$ = positive solution               & $e_{P_1}$ = positive solution               & $e_{P_1}$ = positive solution               & $e_{P_1}$ = negative solution              \\
$e_{P_2}$ = positive solution               & $e_{P_2}$ = negative solution               & $e_{P_2}$ = negative solution               & $e_{P_2}$ = negative solution              \\ 
\multicolumn{2}{c}{Third Quadrant}                                                            & \multicolumn{2}{c}{Fourth Quadrant}                                                          \\
if $\alpha_2$ \textless{}= $-\pi/2$ & if $\alpha_2$ \textgreater \, $-\pi/2$ & if $\alpha_1$ \textless{}= $-\pi/2$ & if $\alpha_1$ \textgreater \, $-\pi/2$ \\
$e_{P_1}$ = negative solution               & $e_{P_1}$ = negative solution               & $e_{P_1}$ = negative solution               & $e_{P_1}$ = positive solution              \\
$e_{P_2}$ = negative solution               & $e_{P_2}$ = positive solution               & $e_{P_2}$ = positive solution               & $e_{P_2}$ = positive solution          \\ \hline \hline   
\end{tabular}
\end{table}
\end{centering}

Once the values of $n_{P_1}$ and $n_{P_2}$ are obtained it is possible to compute the remaining coordinates.	
\begin{equation}
\label{Eq_30}
\begin{gathered}
u_{P_1} = m_{P_1}e_{P_1} - m_{P_1}e_{s/c} + u_{s/c} \hspace{1 cm}
u_{P_2} = m_{P_2}e_{P_2} - m_{P_2}e_{s/c} + u_{sc}
\end{gathered}
\end{equation}	

At this stage the position vectors of the two points $P_1$ and $P_2$ are identified. 
\begin{equation}
\label{Eq_31}
\begin{gathered}
\mathbf{r}_{P_1} = \left[e_{P_1},u_{P_1},0\right] \hspace{2 cm}
\mathbf{r}_{P_2} = \left[e_{P_2},u_{P_2},0\right]
\end{gathered}
\end{equation}	

The ground range angle is simply the angle between these two vectors and can be computed as:	
\begin{equation}
\cos(\Lambda\textsubscript{FOV}) = \frac{\mathbf{r}_{P_1}\cdot\mathbf{r}_{P_2}}{{r}_{P_1}{r}_{P_2}}
\label{Eq_32}
\end{equation}

The last step is to link the aperture angle, $\eta$, with the elevation angle, $\varepsilon$. It should be emphasized that, in this case, there is no close relationship between the aperture angle and the elevation angle, and so the two quantities are independent.
First of all, it is possible to compute the slopes of the tangents to the ellipse in the two points just determined deriving the equations of the two halves of the ellipse. 	
\begin{equation}
\begin{gathered}
m_{t_{P1}} = \pm\frac{\tilde{b} e_{P_1}}{\tilde{a}^2\sqrt{1 - \left(\frac{\scalebox{1}{$e_{P_1}$}}{\scalebox{1}{$\hspace{-0.1 cm}\tilde{a}$}}\right)^2}} \hspace{1 cm}
m_{t_{P2}} = \pm\frac{\tilde{b}e_{P_2}}{\tilde{a}^2\sqrt{1 - \left(\frac{\scalebox{1}{$e_{P_2}$}}{\scalebox{1}{$\hspace{-0.1 cm}\tilde{a}$}}\right)^2}}
\end{gathered}
\label{Eq_33}
\end{equation}	

It needs to be stressed that the ellipse is not a bijective function, and therefore the derivative cannot be considered. The ellipse has to be split in two halves and this is the reason why there are two possible solutions for the slope of the tangents.
The equations of the tangents at the points are known because there are one point and the slope available for each line. Remembering that the elevation angle is the angle between the satellite's center of mass and the observer’s (ground station’s) horizon plane \cite{Cakaj}, it is possible to apply the formulation to compute the angle between two lines without ambiguity because the elevation angle is always lower than $\pi$:	
\begin{equation}
\label{Eq_34}
\begin{gathered}
\tan(\pi - \varepsilon_1) = \frac{m_{t_{P_{1}}} - m_{P_1}}{1 + m_{t_{P_{1}}}m_{P_1}} \hspace{1 cm}
\tan(\pi - \varepsilon_2) = \frac{m_{t_{P_{2}}} - m_{P_2}}{1 + m_{t_{P_{2}}}m_{P_2}}
\end{gathered}
\end{equation}

Of course, the two values of the elevation angles, as for the two values of ground range angles, are different in the ellipsoidal approach because the curvature of the Earth is not constant but changes.
To be consistent with the previous methods that introduce a visibility function depending on the minimum elevation angle for a satellite to be visible, it is useful to express the new approach also in terms of the minimum elevation angle. Basically, there are two solutions that can be adopted for the resolution of this problem: iterative and analytical.

The iterative procedure starts from the value of the half-aperture angle related to the theoretical horizon, for which the elevation angle is zero, and computes the elevation angles for both the two sides giving small decrement to the half-aperture angle. The iteration ends when the difference between the minimum elevation angle and the actual one is below a prescribed threshold. The number of iterations is very small because the values of the minimum elevation angles are in around 5$^{\circ}$-10$^{\circ}$ and the elevation angle is very sensitive to a small variation of the half-aperture angle due to the curvature of the oblate ellipsoid. 

The analytical approach gets the elevation angle as a function of the half-aperture angle in such a way to use directly the right value to determine the coverage area. Starting from Eq. (\ref{Eq_34}), the expressions $m_{t_P}$ is replaced by Eq. (\ref{Eq_33}) obtaining the following intermediate result:

\begin{equation}
-\tan(\varepsilon) = \frac{\pm\tilde{b}e_P-{\tilde{a}}^2m_P\sqrt{1-\left(\frac{\scalebox{1}{$e_P$}}{\scalebox{1}{\hspace{-0.1 cm}$\tilde{a}$}}\right)^2}}{{\tilde{a}}^2\sqrt{1-\left(\frac{\scalebox{1}{$e_P$}}{\scalebox{1}{\hspace{-0.1 cm}$\tilde{a}$}}\right)^2} \pm \tilde{b}e_P}
\label{Eq_35}
\end{equation}
where the identity $\tan(\pi - \varepsilon) = -\tan(\varepsilon)$ has been used. The "+" sign is related to the negative half part of the ellipse, whereas the "-" is associated to the positive one.
Finally, by replacing $e_P$ with Eq. (\ref{Eq_29}) the only unknown in the equations is the angular coefficient $m_P$, which is directly associated to the half-aperture angle $\eta$. The advantage of the last method is to have an analytical relation between the elevation angle and the half-aperture angle. However, the equation is highly non-linear, is transcendental, and needs to be solved numerically paying attention to the correct initial guess or interval used.

\section{Determination of the Coverage Region in Relevant Satellite Orbital Scenarios}

In this section the field of view of a generic satellite is computed in different orbital scenarios. Indeed, during the preliminary mission analysis according to the mission requirements, the attitude of the satellite can be chosen among three different pointings: a geocentric pointing, where the line of sight of the navigation antenna is directed toward the Earth center; a geodetic pointing allowing the direction of the line of sight to be normal to the local horizon on the surface of the Earth; and a generic pointing associated to a casual direction to analyze the effects of the attitude perturbations.

\subsection{Generic Pointing Case}

The first scenario is most general one involving a moving line of sight. From this analysis, all the simplified cases may be retrieved by imposing the right value of the geometric parameters.

To apply this procedure, it is assumed that the position of the satellite's center of mass is known together with the direction cosines of the navigation antenna line of sight that can be the output of an attitude and orbit propagator. It is possible to work both with the spacecraft position vector and direction cosines of the navigation antenna  expressed in the inertial reference frame and relative to a fixed Earth. The output of the procedure will result in position vectors consistent with the type of representation used. The first thing to be done is the determination of the projection of the line of sight onto the Earth's surface corresponding to the nadir. Indeed, if the line of sight is not aligned with the geocentric direction also the footprint changes. It is useful to express the equation of the line of sight in its parametric form considering one point (the spacecraft center of mass) and the direction cosines:
\begin{equation}
\begin{cases}
x(t) = x_{s/c} + n_1t \\
y(t) = y_{s/c} + n_2t \\
z(t) = z_{s/c} + n_3t
\end{cases}
\label{Eq_36}
\end{equation}

Because the footprint belong to the Earth's surface, its coordinates have to satisfy the equation of the oblate ellipsoid:
\begin{equation}
	\frac{(x_{s/c} + n_1t)^2 + (y_{s/c} + n_2t)^2}{R^2_{eq}} + \frac{(z_{s/c} + n_3t)^2}{R^2_{pol}} = 1
	\label{Eq_37}
\end{equation}
where $R$\textsubscript{eq} and $R$\textsubscript{pol} are the Earth equatorial and polar radii, respectively.
After performing some operations, the equation may be rearranged in the following form:
\begin{equation*}
	\scalebox{0.94}{$(n^2_1R^2_{pol}+n^2_2R^2_{pol} + n^2_3R^2_{eq})t^2 + 2(n_1x_{s/c}R^2_{pol} + n_2y_{s/c}R^2_{pol} + n_3z_{s/c}R^2_{eq})t + (x^2_{s/c}R^2_{pol} + y^2_{s/c}R^2_{pol} + z^2_{s/c}R^2_{eq} - R^2_{eq}R^2_{pol}) = 0$}
\end{equation*}

\begin{figure}[ht!]
	\centering
	\includegraphics[scale = 0.6]{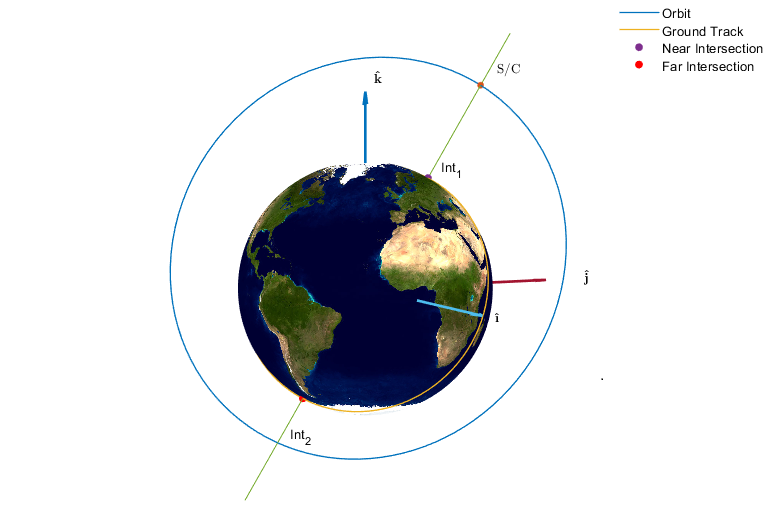}
	\caption{Intersection points of a generic line of sight with the oblate ellipsoid of rotation.}
	\label{Fig_5}
\end{figure}

Two solutions are obtained for the parameter $t$:
\begin{equation}
\label{Eq_38}
	t_{1,2} = \frac{-(n_1x_{s/c}R^2_{pol} + n_2y_{s/c}R^2_{pol} + n_3z_{s/c}R^2_{eq}) \pm \sqrt{\Delta_{int}}}{(n^2_1R^2_{pol} + n^2_2R^2_{pol} + n^2_3R^2_{eq})}
\end{equation}
where $\Delta$\textsubscript{int} is the discriminant of the equation and it is equal to:
\begin{align*}
	\Delta_{int} = 2n_1n_2x_{s/c}y_{s/c}R^4_{pol} + 2n_1n_3x_{s/c}z_{s/c}R^2_{eq}R^2_{pol} + 2n_2n_3y_{s/c}z_{s/c}R^2_{eq}R^2_{pol} - n^2_1y^2_{s/c}R^4_{pol} - n^2_1z^2_{s/c}R^2_{eq}R^2_{pol}+ \\+ n^2_1R^2_{eq}R^4_{pol} - n^2_2x^2_{s/c}R^4_{pol} - n^2_2z^2_{s/c}R^2_{eq}R^2_{pol} + n^2_2R^2_{eq}R^4_{pol} - n^2_2x^2_{s/c}R^2_{eq}R^2_{pol} - n^2_3y^2_{s/c}R^2_{eq}R^2_{pol} + n^2_3R^4_{eq}R^2_{pol}
\end{align*}
There are two solutions because this method considers both the entry point and the exit point of the line of sight from the ellipsoid as shown in Fig. \ref{Fig_5}. 

Because in the parametric equation of the line of sight the coordinates of the spacecraft are fixed at a given time instant and the direction cosines are related to the pointing toward the Earth's center, it happens that the greater the magnitude of the parameter $t$, the greater the distance of the computed point from the satellite's center of mass is. So, the exact solution is the one associated to the minimum value of the parameter $t$. 

\begin{figure}[ht!]
\centering
\begin{subfigure}{.5\textwidth}
\centering
   \includegraphics[scale = 0.45]{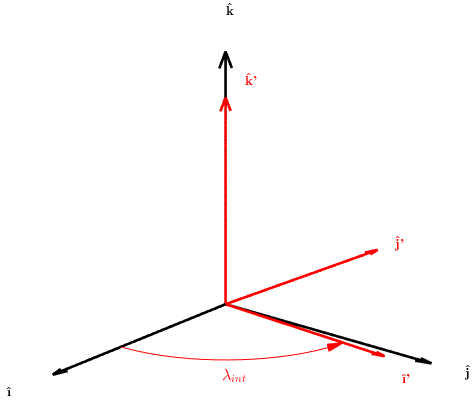}
  \caption{Longitude rotation}
  \label{Fig_6a}
\end{subfigure}%
\begin{subfigure}{.5\textwidth}
\centering
  \includegraphics[scale = 0.5]{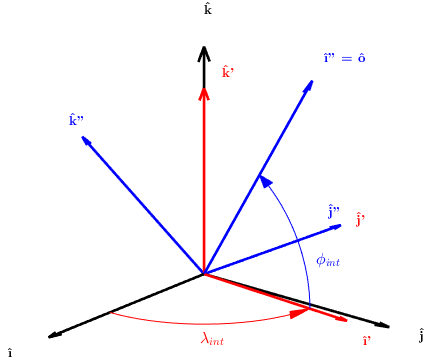}
  \caption{Latitude rotation}
  \label{Fig_6b}
\end{subfigure}
\caption{First series of rotations.}
\label{Fig_6}
\end{figure}

The next step is to transform the 3D representation of the problem into a planar one in order to use the techniques developed in the previous section. For this reason, more than one rotation matrix should be introduced in order to express the coordinates of the spacecraft into the local reference frame defined before.
The first series of rotations allows to align the $\mathbf{\hat{i}}$ axis of the inertial frame with the direction of the line of sight as shown in Fig. \ref{Fig_6}:

\begin{enumerate}
\item A first positive rotation around the $\mathbf{\hat{k}}$-axis of an angle equal to the longitude associated to the line of sight. 
\item A second negative rotation around the $\mathbf{\hat{j'}}$-axis of an angle equal to the latitude associated to line of sight.	
\end{enumerate} 

It is easy to compute the longitude and the latitude of the line of sight because the direction cosines are known. This operation has to be done considering the same direction but the opposite pointing. Indeed, the line of sight is directed from the spacecraft toward the Earth and it is necessary to introduce the opposite direction $\mathbf{\hat{o}}$.
\begin{equation}
	\mathbf{\hat{o}} = -\mathbf{\hat{n}}
	\label{Eq_39}
\end{equation}

By applying the conversion from Cartesian coordinates to spherical coordinates, the result is the following:
\begin{equation}
	\begin{gathered}
	\lambda_{int} = {\tan}^{-1}\left(\frac{o_2}{o_1}\right) \\
	\phi_{int} = {\sin}^{-1}(o_3)	
	\end{gathered}
	\label{Eq_40}
\end{equation} 
with $\lambda_{int}$ and $\phi_{int}$ the longitude (in-plane) and latitude (elevation) angles associated to the line of sight. Once the two angles have been derived, the two rotation matrices can be written:
\begin{equation}
\label{Eq_41}
	\mathbf{R_1} = \begin{bmatrix}
	\cos(\lambda_{int}) & \sin(\lambda_{int}) & 0 \\
	-\sin(\lambda_{int}) & \cos(\lambda_{int}) & 0 \\
	0              &           0          & 1 
	\end{bmatrix}	
	\hspace{2 cm}
	\mathbf{R_2} = \begin{bmatrix}
	\cos(\phi_{int}) & 0 & \sin(\phi_{int}) \\
	0 & 1 & 0 \\
	-\sin(\phi_{int}) &  0  & \cos(\phi_{int}) 
	\end{bmatrix}
\end{equation}
that aligns the $\mathbf{\hat{i}}$ axis of the geocentric inertial frame with the line of sight of the navigation antenna. 

The projection of the conical field of view onto the Earth's surface cannot be expressed using a planar function if the ellipsoidal model is adopted. Because it is difficult to derive an analytical expression for such three-dimensional surface, it is more convenient to define it in terms of couples of points lying on this surface computed with Eqs. (\ref{Eq_29}) and (\ref{Eq_30}).
For this reason, all the planes generated by the rotation of the $i''j''$ plane around the line of sight, corresponding to the axis of the new reference frame, have to be considered and for each of them a couple of points are derived. 
This means that a new rotation matrix of an angle $\psi$, which is variable in the range $[0,\pi]$, needs to be introduced:
\begin{equation}
	\mathbf{R_3} = \begin{bmatrix}
	1 & 0 & 0 \\
	0 & \cos(\psi) & \sin(\psi) \\
	0 &  -\sin(\psi) & \cos(\psi) 
	\end{bmatrix}
	\label{Eq_42}
\end{equation} 

The final rotation matrix will be a combination "321" with the following form:
\begin{equation}
		\scalebox{0.87}{$\mathbf{A_{321}} = 
		\begin{bmatrix} \cos(\phi_{int})\cos(\lambda_{int}) & \cos(\phi_{int})\sin(\lambda_{int}) & \sin(\phi_{int}) \\
		-\cos(\psi)\sin(\lambda_{int})-\sin(\psi)\sin(\phi_{int})\cos(\lambda_{int}) & \cos(\psi)\cos(\lambda_{int}) - \sin(\psi)\sin(\phi_{int})\sin(\lambda_{int}) & \sin(\psi)\cos(\phi_{int}) \\
		\sin(\psi)\sin(\lambda_{int}) - \sin(\phi_{int})\cos(\lambda_{int})\cos(\psi) &  -\sin(\psi)\cos(\lambda_{int}) - \sin(\phi_{int})\sin(\lambda_{int})\cos(\psi) & \cos(\psi) \cos(\phi_{int})
		\end{bmatrix}$}
        \label{Eq_43}
\end{equation} 

\begin{figure}[ht!]
	\centering
	\includegraphics[scale = 0.64]{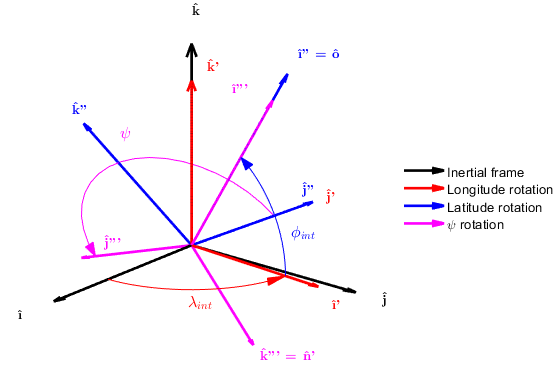}
	\caption{Transformation from the geocentric inertial frame to the generic frame aligned with the line of sight.}
	\label{Fig_7}
\end{figure}

A plane in the space is unequivocally identified if the direction cosines of a line normal to the plane itself and a point lying on the plane are known because it is necessary to compute the distance $d$ of the plane from the origin of the reference system. After the rotation of the reference system, the normal to each plane is simply identified by the unit vector $\mathbf{\hat{n}'}=[0,0,1]$, expressed in the rotated reference frame, as shown in Fig. \ref{Fig_7}.  
This normal must be projected in the inertial reference frame and for this reason the inverse rotation matrix needs to be defined. One of the properties of the rotation matrices is that the inverse operation is equivalent to the transposition of the matrix, and so it is possible to write:
\begin{equation}
	\mathbf{\hat{n}'} = \mathbf{A^T_{321}}\left[0,  0,  1\right]^{\mathbf{T}}
	\label{Eq_44}
\end{equation} 

The other information needed to write the analytical expression of the plane is also known since the center of mass of the spacecraft lies on all the planes being part of the axis of rotation around which each plane is obtained. For this reason the distance $d$ of each plane from the origin of the inertial reference system is:
\begin{equation}
	d = n'_1x_{s/c} + n'_2y_{s/c} + n'_3z_{s/c}
	\label{Eq_45}
\end{equation}
where $[x_{s/c},y_{s/c},z_{s/c}]$ represent the coordinates of the spacecraft position vector in the inertial frame. If a Keplerian motion is assumed for the spacecraft, the spacecraft position vector in the inertial frame can be derived from the knowledge of its orbital parameters. Following Bate et al. \cite{Bate}:

\begin{equation}
\label{Eq_46}
\begin{bmatrix}
	x_{s/c} \\
	y_{s/c} \\
	z_{s/c} 
	\end{bmatrix} = r_{s/c}
	\begin{bmatrix}
    \cos(\Omega)\cos(\nu) - \sin(\Omega)\cos(i)\sin(\nu) \\
     \sin(\Omega)\cos(\nu) + \cos(\Omega)\cos(i)\sin(\nu)\\
    \sin(i)\sin(\nu)
	\end{bmatrix}
\end{equation}

The last step is to align the $i''',j''',k'''$ reference system with the local reference system of the ellipse $e,u,n'$. Indeed, in the general case, the line of sight do not pass through the center of the ellipse but it is coplanar with the apse line direction. Therefore, once the apse line direction in the inertial frame is obtained with Eq. (\ref{Eq_11}), the angle between the line of sight and the apse line axis is computed:

\begin{equation}
	\cos(\beta) = \frac{\mathbf{r}\textsubscript{line}\cdot\mathbf{\hat{e}}}{{r}\textsubscript{line}}
	\label{Eq_47}
\end{equation} 

\begin{figure}[ht!]
	\centering
	\includegraphics[scale = 0.85]{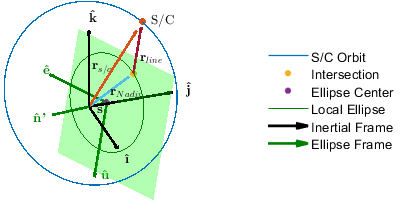}
	\caption{Representation of the local reference frame and the local ellipse.}
	\label{Fig_8}
\end{figure}

The $\mathbf{r}$\textsubscript{line} is the vector connecting the nadir and the satellite's center of mass as depicted in Fig. \ref{Fig_8}. This vector can be expressed in the inertial frame thanks to the relative law:

\begin{equation}
	\mathbf{r}\textsubscript{line} = \mathbf{r}_{s/c} - \mathbf{r}\textsubscript{Nadir}
	\label{Eq_48}
\end{equation}

\begin{figure}[ht!]
	\centering
	\includegraphics[scale = 0.75]{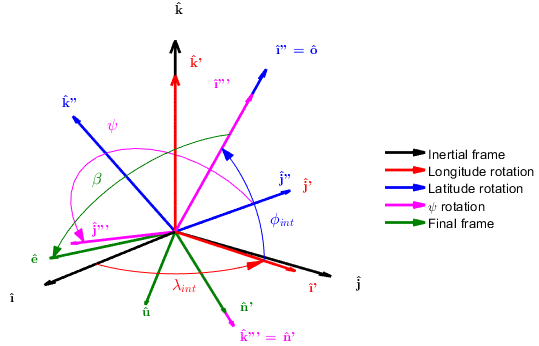}
	\caption{Transformation from the geocentric inertial frame to the local ellipse reference frame.}
	\label{Fig_9}
\end{figure}

It is convenient to express the previous quantities in the $i''',j''',k'''$ reference frame aligned with the line of sight. Indeed, in this case it is possible to identify if the angle between the two directions is associated to a clockwise or counterclockwise rotation.
\noindent
Another rotation matrix around the $\mathbf{\hat{n}'}$-axis which is coincident with the normal to the plane of the intersected ellipse is computed:

\begin{equation}
\label{Eq_49}
	\mathbf{R_{\beta}} =	\begin{bmatrix}
			\cos(\beta) & \sin(\beta) & 0 \\
			-\sin(\beta) & \cos(\beta) & 0 \\
			0              &           0          & 1 
		\end{bmatrix}
\end{equation}
If the $\mathbf{A}$\textsubscript{321} matrix previously computed in Eq. (\ref{Eq_43}) is left multiplied for the $\mathbf{R_{\beta}}$ matrix, the final rotation matrix from the inertial reference frame to the local reference frame of the ellipse is obtained as shown in Fig. \ref{Fig_9}.

Another way to get the final rotation matrix from the geocentric reference system to the local reference system $\mathbf{\hat{e}},\mathbf{\hat{u}},\mathbf{\hat{n}'}$ is to remember the definition of rotation matrix as the projections of the unit vectors of the rotated reference frame onto the initial reference frame. This is easily done once the position vectors of the rotated reference frame are expressed in the inertial reference frame. By using Eqs. (\ref{Eq_11}), (\ref{Eq_13}), and (\ref{Eq_44}) to derive, respectively, the unit vectors $\mathbf{\hat{e}},\mathbf{\hat{u}},\mathbf{\hat{n}'}$ in the inertial reference frame, the matrix $\mathbf{A}\textsubscript{rot}$ is:

\begin{equation}
\label{Eq_50}
\mathbf{A}\textsubscript{rot} = 
\begin{bmatrix}
e_x & e_y & e_z \\
u_x & u_y & u_z \\
{n'}_x & {n'}_y & {n'}_z
\end{bmatrix}
\end{equation}

The last passage is to convert the coordinates of the satellite's center of mass from the inertial to the local frame:

\begin{equation}
\label{Eq_51}
	\begin{bmatrix}
	e_{s/c} \\
	u_{s/c} \\
	0 
	\end{bmatrix} = \mathbf{A}\textsubscript{rot}
	\begin{bmatrix}
    x_{s/c} \\
    y_{s/c} \\
    z_{s/c}
	\end{bmatrix}	
	\hspace{1 cm} \textrm {with} \hspace{0.5 cm} \mathbf{A}\textsubscript{rot} = \mathbf{R_{\beta}}\mathbf{A}\textsubscript{321}
\end{equation}

Everything is ready to apply the technique analyzed in Sec. III and to get the intersection points $P_1$ and $P_2$.
The intersection points are defined in the local reference system. Therefore, with the inverse rotation matrix the points are expressed in the inertial reference frame.
\begin{equation}
\label{Eq_52}
	\begin{gathered}
	\mathbf{r}_{P_{1\textsubscript{relative}}} = \mathbf{A^T_{\textnormal{rot}}}\mathbf{r}_{P_{1\textsubscript{local}}} \\
	\mathbf{r}_{P_{2\textsubscript{relative}}} = \mathbf{A^T_{\textnormal{rot}}}\mathbf{r}_{P_{2\textsubscript{local}}}
	\end{gathered}	
\end{equation}

The procedure is not completed because the center of the local reference frame is not coincident with the center of the inertial reference frame. This means that an additional operation must be performed to obtain the right points. By applying the laws of the relative reference systems, we obtain:

\begin{equation}
\label{Eq_53}
	\begin{gathered}
	\mathbf{r}_{P_{1\textsubscript{inertial}}} = \mathbf{s} + \mathbf{r}_{P_{1\textsubscript{relative}}} \\
	\mathbf{r}_{P_{2\textsubscript{inertial}}} = \mathbf{s} + \mathbf{r}_{P_{2\textsubscript{relative}}} 
	\end{gathered}
\end{equation}

\subsection{Geodetic Pointing}

\begin{figure}[ht!]
	\centering
	\includegraphics[scale = 0.8]{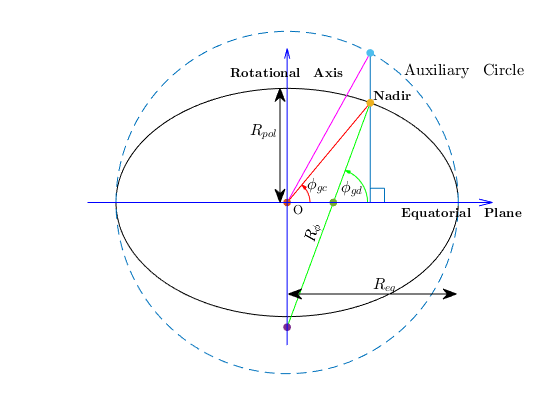}
	\caption{Geodetic latitude representation.}
	\label{Fig_10}
\end{figure}

While for the spherical approach there is no difference between geocentric and geodetic pointing, for the ellipsoidal approach this difference is present due to the geometry considered. Indeed, in an oblate ellipsoid of rotation the conjunction of a point with the center is different from the local vertical of that point to the surface as shown in Fig. \ref{Fig_10}. However, some some simplifications can be performed to the generic pointing to derive the same results.

The first thing to be emphasized is that, also in this case, the local reference frame is not centered in the origin of the inertial reference system. The nadir formulation can be reformulated in a lighter form considering the properties of the ellipsoid. Following Curtis \cite{Curtis}:
\begin{equation}
\label{Eq_54}
\begin{bmatrix}
x_{int} \\
y_{int} \\
z_{int} 
\end{bmatrix} = 
 \begin{bmatrix}
(R_{\phi} + H)\cos(\phi_{gd})\cos(\theta) \\
(R_{\phi} + H)\cos(\phi_{gd})\sin(\theta)
\\
(1-f^2)(R_{\phi} + H)\sin(\phi_{gd})
\end{bmatrix}
\hspace{2 cm}
R_{\phi} = \frac{R_{eq}}{\sqrt{1 - (2f - f^2)\sin^2(\phi_{gd})}}
\end{equation}
where $\theta$ is the local sideral time, $H$ is the height of the satellite and $f$ represents the flattening of the Earth equal to 0.00335 according to the WGS84 model representation of the Earth \cite{WGS84}.
The relative position vector of the spacecraft with respect to the nadir may be evaluated in the inertial reference frame. The method proceeds in a similar way like the generic pointing with the only difference that in the determination of the $\mathbf{A}$\textsubscript{321} matrix the geodetic latitude is used.

\subsection{Geocentric Pointing Case}

The geocentric pointing case can be retrieved from the generic pointing approach imposing the value of the direction cosines equal to the direction cosines of the satellite's position vector.

\section{Numerical Simulations}

In Sec. III the formulas to compute the intersection of a conical field of view with the Earth modeled as an oblate ellipsoid of rotation have been derived. The same formulas have been applied to compute the coverage area for different types of pointing.
The first step is to verify that the points computed with the new approach are actually on the Earth's surface modeled as oblate ellipsoid of rotation.
Each point that lies on the surface of the ellipsoid shall fulfill the equation:

\begin{equation}
\label{Eq_55}
	\frac{x^2 + y^2}{R^2_{eq}} + \frac{z^2}{R^2_{pol}} = 1
\end{equation}
where $R$\textsubscript{eq} and $R$\textsubscript{pol} are the Earth equatorial and polar radii and their numerical values are $6378.137$ km and $6356.752$ km, respectively, according to the World Geodetic System (WGS) 1984 \cite{WGS84}.
It is possible to compute the value obtained substituting the coordinates of each point obtained from the formulas and compare with respect to the right-hand side equal to 1. This operation can be performed considering the three types of pointing.
After several simulations considering also different values of the half-aperture angle, $\eta$, for a geocentric pointing the maximum error experienced is equal to $9\cdot10^{-7}$ km, whereas for a geodetic pointing the error decreases to $9\cdot10^{-8}$ km. The maximum error is still the same for a generic pointing where the line of sight, and so the conical field of view, is not fixed but it is moving. 

After verifying the formulations derived, another step is to show that the intersection of the conical field of view with the oblate ellipsoid is not a planar line, but a three-dimensional surface. This operation can be done considering a simple linear algebra operation. Indeed, considering four generic points belonging to the coverage area, it is possible to prove if they are coplanar computing the following determinant:

\begin{equation}
\label{Eq_56}
	\det\begin{bmatrix}
	x_3 - x_0 & y_3 - y_0 & z_3 - z_0 \\
	x_2 - x_0 & y_2 - y_0 & z_2 - z_0 \\
	x_1 - x_0 & y_1 - y_0 & z_1 - z_0  
	\end{bmatrix}
\end{equation}

If the value of the determinant is equal to zero, it means that the four points are coplanar. Also in this case, after several simulations the value of the determinant is always finite and different from zero. Therefore, it is not possible to express the intersection of a conical field of view with an oblate ellipsoid as a function of a single variable. 
\begin{table}[!h]
\centering
\caption{Procedure for the computation of the coverage area}
\label{Tab_2}
\begin{tabular}{lc}
\hline \hline
Input   &  Satellite Position $\mathbf{r}_{s/c}$, line of sight direction $\mathbf{\hat{n}}$     \\ 
Step 1  & Line-of-sight intersection with the equatorial plane $\mathbf{P}$\textsubscript{int} [Eqs. (\ref{Eq_36}) and (\ref{Eq_38})]                      \\ 
Step 2  & Determination of the geographic longitude $\lambda$\textsubscript{int} and latitude $\phi$\textsubscript{int} [Eq. (\ref{Eq_40})]                           \\ 
Step 3  & Alignment of the $\mathbf{\hat{I}}$ axis of the geocentric frame with the line-of-sight direction $\mathbf{\hat{o}}$ [Eq. (\ref{Eq_41})] \\ 
Step 4  & Intersection of the plane normal to $\mathbf{\hat{n}'}$ with the oblate ellipsoid of rotation     \\ 
Step 5  & Computation of the geometric parameters of the resulting ellipse [Eqs. (\ref{Eq_9}-\ref{Eq_13})]                \\ 
Step 6  & Alignment of the $\mathbf{\hat{I}}$ axis of the geocentric frame with apse line direction $\mathbf{\hat{e}}$       \\ 
Step 7  & Computation of the two ellipse points belonging to the coverage area [Eqs. (\ref{Eq_29}), (\ref{Eq_30}), (\ref{Eq_50}), and (\ref{Eq_51})]             \\ 
Step 8  & Rotation of the plane normal to $\mathbf{\hat{n}'}$ around the line-of-sight direction $\mathbf{\hat{o}}$ [Eq. (\ref{Eq_43})]     \\ 
Step 9  & New intersection with the oblate ellipsoid of rotation                           \\ 
Step 10 & Derivation of group of points belonging to the coverage area                     \\ 
Step 11 & Interpolation of the points to get the analytical formulation of the surface     \\ 
Step 12 & Identification of the points inside and outside the coverage area           \\ \hline \hline 
\end{tabular}
\end{table} 

This is the first main difference with respect to modeling the Earth shape as a sphere. 
The only solution to get an analytical formulation of the previous three-dimensional surface is to rely on an interpolation that introduces an approximation. The interpolation procedure is more refined as the number of points used to perform the interpolation increases.
However, it is possible to compute as many points as needed using the procedure described in Sec. III. The problem is solved using a function which takes in input a series of points of the same generic dimension, and the coefficients of the independent variables desired as output of the function to express the interpolation. This function is named \textit{Polyfitn} \cite{DErrico} and is available on the MathWorks website.
For the determination of the analytical expression of the three-dimensional surface only the odd coefficients of the independent variables (i.e., $"x, y, x^3, y^3"$) are considered in such a way to distinguish in a unique way the right octant. Table \ref{Tab_2} summarizes the complete procedure for the computation of the coverage area starting from the position of the satellite and the half-aperture angle $\eta$ of the navigation antenna. 

\subsection{Sphere Versus Oblate Ellipsoid}

It is important to compare the results obtained considering the Earth as a sphere with the ones obtained considering the oblate Earth. The aim of the analysis is to prove that the spherical model is a first-order model that can be used whenever the accuracy required is not very tight.
The error between two identical quantities evaluated with the two different models is computed and the variation of this error versus relevant parameters such as the eccentricity, the inclination of the orbit, and the half-aperture angle is analyzed starting from a Galileo-like satellite orbit as reference whose parameters are given in Table \ref{Tab_3}.

\begin{table}[hbt!]
\caption{\label{tab:table1} Nominal parameters used for the numerical simulations}
\centering
\begin{tabular}{lcccccc}
\hline \hline
Height, km&  Eccentricity & Inclination, deg & RAAN, deg & Pericenter anomaly, deg & $\eta$, deg& \\\hline
23,229.32& 0& 56& 0& 0& 10\\
\hline \hline
\end{tabular}
\label{Tab_3}
\end{table}

\begin{figure}[ht!]
	\centering
	\includegraphics[scale = 0.65]{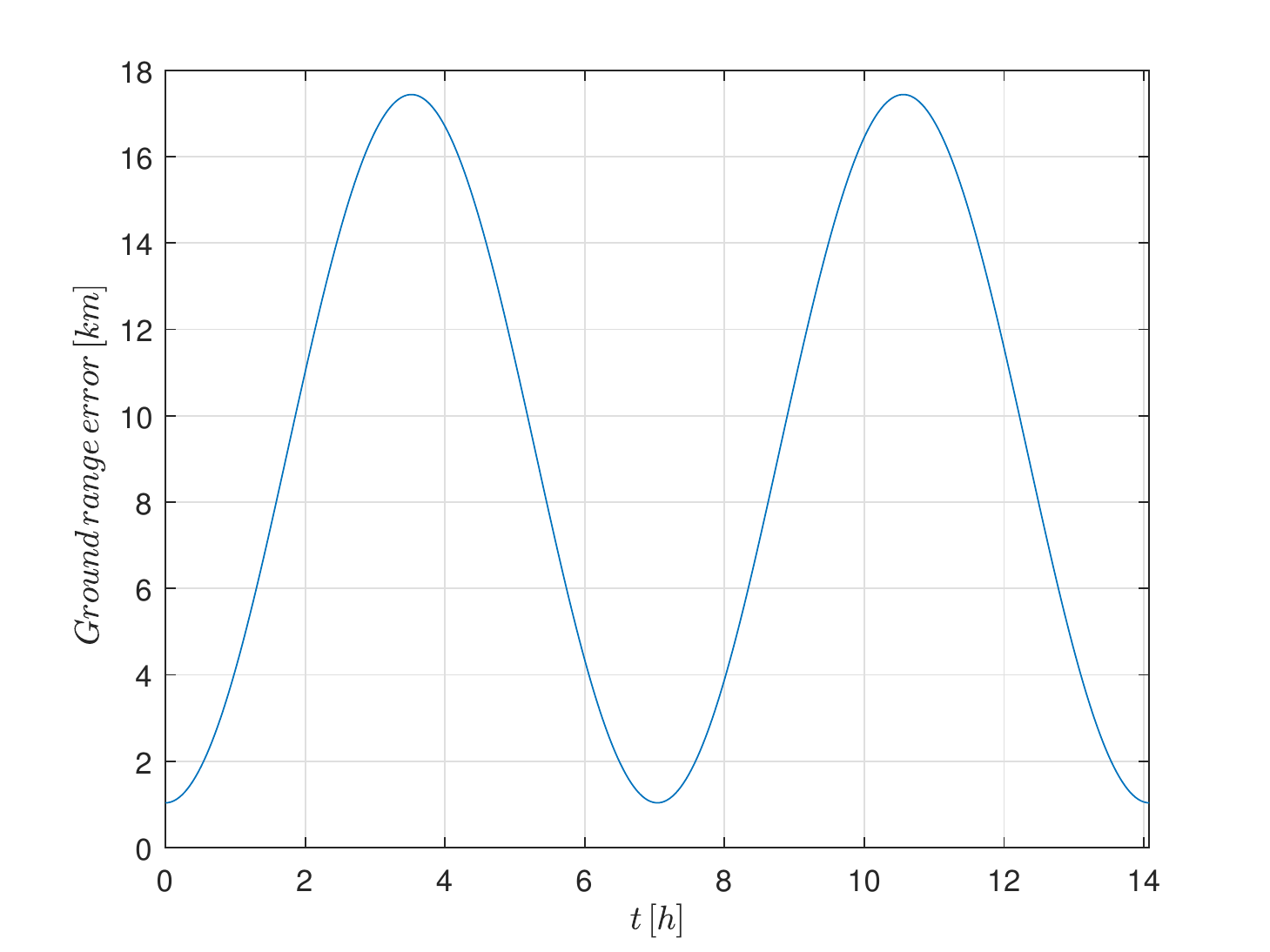}
	\caption{Ground range difference evolution in one orbit for geocentric pointing.}
	\label{Fig_11}
\end{figure}

In Fig.\hspace{1.5 mm}\ref{Fig_11} the ground range error computed as the difference between the spherical and the ellipsoidal ground range is plotted for a single revolution of the satellite whose orbital parameters are reported in Table \ref{Tab_3}. The ground range is the distance measured between two points belonging to the coverage area that are symmetrical with respect to the projection of the satellite on the Earth's surface (nadir). Repeating the simulation by changing the orbit shows that the behavior of this variable is the same. This plot is somehow expected and it is an additional proof for the validity of the ellipsoidal model. 
Indeed, the minimum difference is experienced whenever the spacecraft is nearby the equator, whereas the maximum difference is obtained as the spacecraft is moving toward the poles. This happens since the spherical and the ellipsoidal geometry are more similar at the equator and reach the maximum deviation at the poles. 

It is important to emphasize the magnitude of the ground range difference. The maximum deviation is $20$ km and this number can be considered significantly high for many applications, such as navigation mission and GNSS services, where the precision required is on the order of meters. Therefore, the spherical model is not completely wrong but it cannot be used in such accurate applications. 
\begin{figure}[ht!]
 	\centering
 	\includegraphics[scale = 0.65]{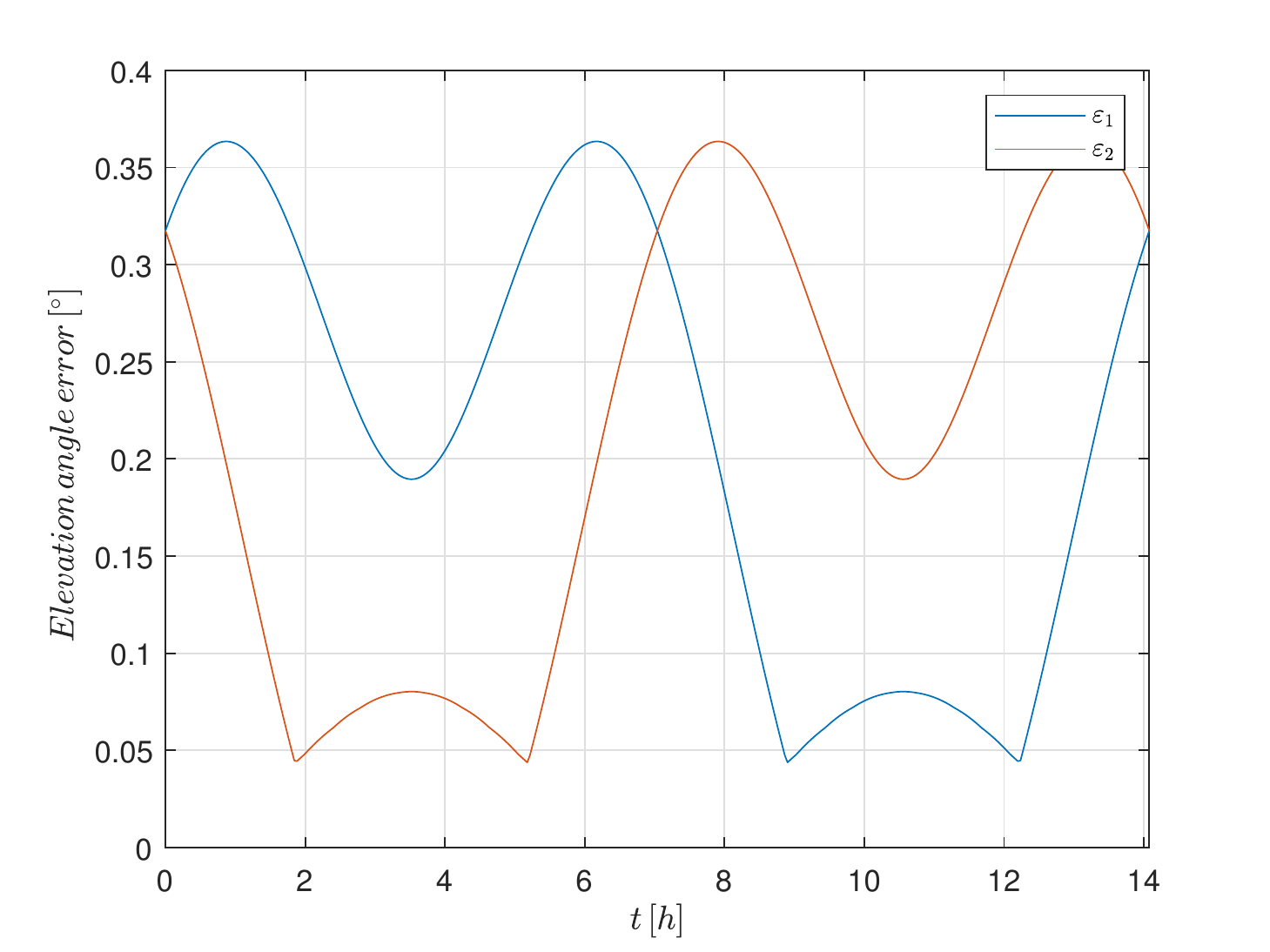}
 	\caption{Elevation angle difference evolution in one orbit for geocentric pointing.}
 	\label{Fig_12}
\end{figure}

Next to the ground range difference, another important variable is the elevation angle because it is used inside the requirements for a navigation satellite to determine the limit of visibility. For a spherical model the elevation angle is strictly related to the half-aperture angle and it is a constant due to the spherical symmetry. This is not valid for the ellipsoidal approach because the geometry is different and the curvature changes according to the position of the spacecraft.
In Fig.\hspace{1.5 mm}\ref{Fig_12} the behavior of the elevation angle error is presented. The error is computed as the difference between the minimum elevation angles associated to a particular conical field of view modeling the Earth's surface as a sphere and as an oblate ellipsoid. In a sphere there is no difference between the two sides of the conical field of view with respect to the line of sight. For an ellipsoid the curvature from one side is different from the other one. The points of intersection of the two lines occurs when the spacecraft is at the equator because it is the only case in which the spacecraft sees a symmetrical geometry with respect to the line of sight. In all the other cases the curvature of the ellipsoid changes continuously. It is important to stress that an error of $0.5^{\circ}$ on the elevation angle is relatively high and is associated to a wide area on the Earth's surface. For this reason there is the possibility that some regions on the Earth's surface are inside the field of view of the satellite or the other way around. 
\begin{figure}[ht!]
	\centering
	\includegraphics[scale = 0.65]{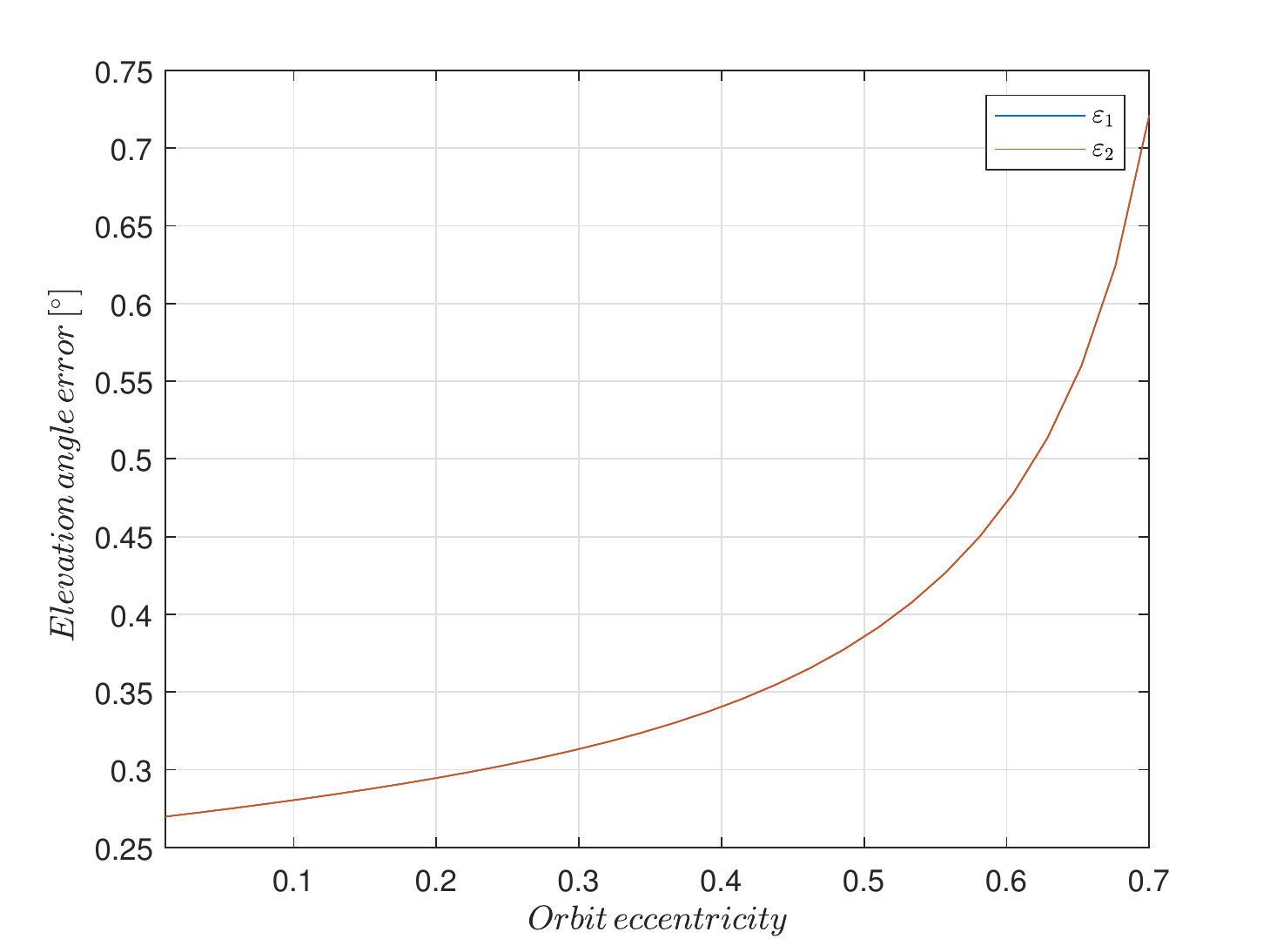}
	\caption{Elevation angle difference vs orbit eccentricity for geocentric pointing.}
	\label{Fig_13}
\end{figure}

\begin{figure}[ht!]
	\centering
	\includegraphics[scale = 0.65]{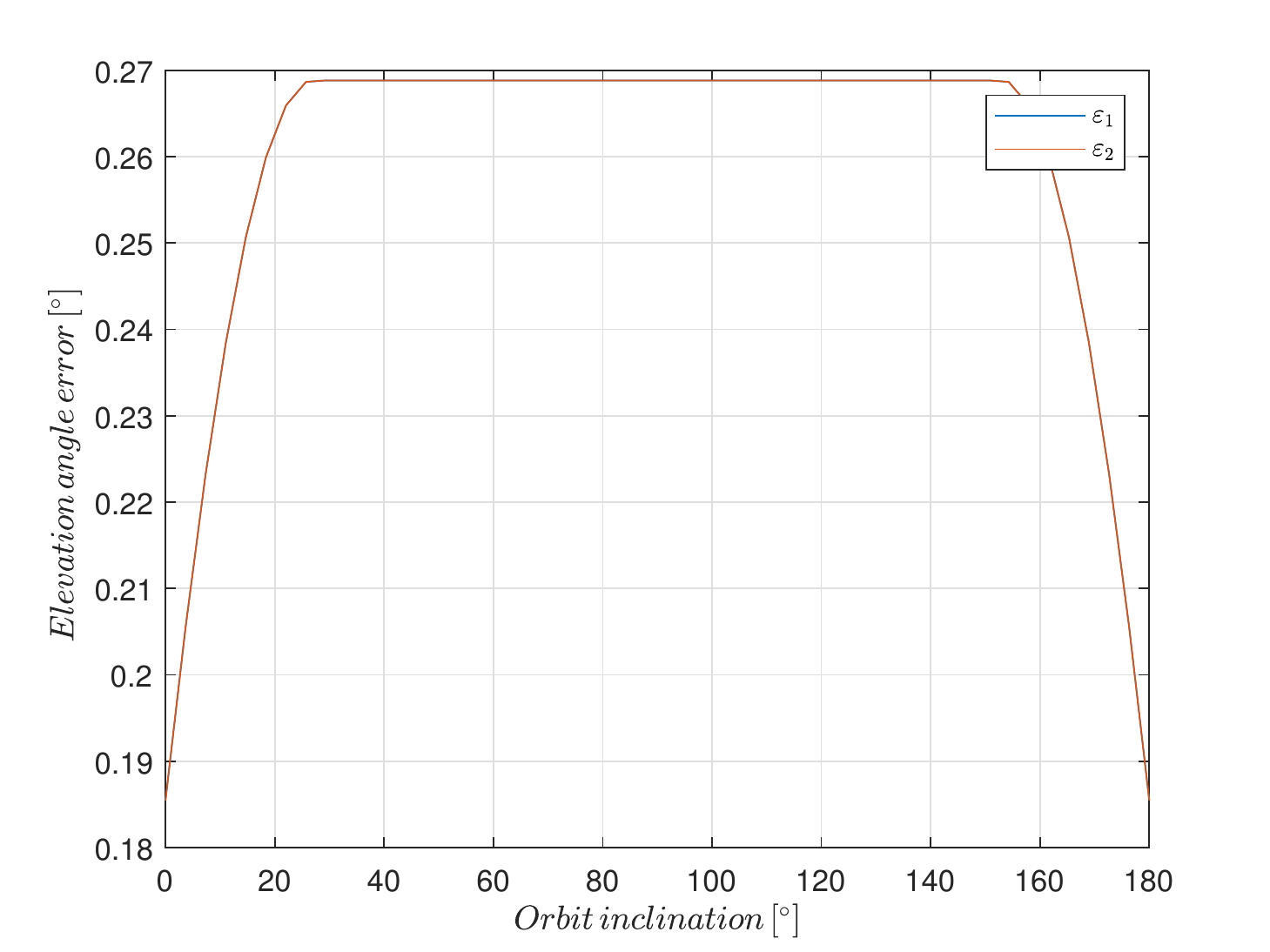}
	\caption{Elevation angle difference vs orbit inclination for geocentric pointing.}
	\label{Fig_14}
\end{figure}

It is also interesting to understand how the elevation angle changes according to the type of orbit.
In Fig.\hspace{1.5 mm}\ref{Fig_13} the two lines are overlapped because the maximum value of the elevation angle error is computed for each orbit and these maximum values are the same for the two elevation angles. 
The elevation angle error does not increase linearly with the eccentricity of the orbit. This is right since the higher the eccentricity, the higher the distance that the spacecraft reaches. If the distance from the Earth increases, the spacecraft can look at a larger region characterized by a greater curvature with respect to the spherical case where there is a change in the elevation angle only linked to the distance of the spacecraft. 
In Fig.\hspace{1.5 mm}\ref{Fig_14} the elevation angle error is showed as a function of the inclination.
This behavior is related to the geometry of the ellipsoid. Indeed, as the inclination of the orbit increases it is easier to look at points belonging to different curvatures of the ellipsoid and to localize the maximum deviations. Note that this is not the elevation angle difference during the orbit but its maximum value for different inclinations. This is the reason why the elevation angle error keeps constant in a particular range of inclination angles.

\subsection{Visibility Period}
The last application involving the proposed approach is the determination of the intervals of viewing for a generic satellite considering prescribed ground stations.
Sentinel-2A satellite has been chosen together with three ground stations to simulate a real scenario:
\begin{enumerate}
    \item ASI Matera Laser Ranging Observatory (MLRO) \cite{Materags}
    \item ESTRACK Maspalomas radio antenna ground station \cite{Maspalomasgs}
    \item KSAT Svalbard satellite station \cite{Svalbardgs}
\end{enumerate}

Table \ref{Tab_4} shows Sentinel-2A osculating orbital elements obtained from the two-line Elements (TLE) set at the epoch 25/02/2019 8:40:17 \cite{SentinelTLE}.

\begin{table}[hbt!]
\caption{Sentinel-2A osculating orbital elements at epoch 25/02/2019 08:40:17}
\centering
\begin{tabular}{cccccc}
\hline \hline
\begin{tabular}[c]{@{}c@{}}Semi-major\\ axis, km\end{tabular} & Eccentricity & Inclination, deg & RAAN, deg & \begin{tabular}[c]{@{}c@{}}Pericenter\\ anomaly, deg\end{tabular} & \begin{tabular}[c]{@{}c@{}}Mean \\ anomaly, deg\end{tabular} \\ \hline
7167.129                                                      & 0.000132     & 98.5657          & 132.4338  & 76.3371                                                           & 238.7960                                                     \\ \hline \hline
\end{tabular}
\label{Tab_4}
\end{table}
\noindent

Sentinel-2A position has been propagated for one solar day under the assumption of Keplerian motion, and the rise/set times with respect to the ground stations have been computed considering a minimum elevation angle of 5$^{\circ}$. This means that the interval of viewing is defined as the time as long as the elevation angle of the ground station with respect to the satellite is higher than 5$^{\circ}$.
Table \ref{Tab_5} summarizes the results of the numerical simulations presenting the geographic coordinates of the three ground stations according to the WGS84 \cite{WGS84} ellipsoidal model of Earth and the rise/set times associated to each passage of the satellite for each ground station.

\begin{table}[h!]
\caption{Rise/set times for Sentinel-2A}
\centering
\begin{tabular}{cccccc}
\hline \hline
Ground station & Longitude, deg & Latitude, deg & Height, m & Rise times, s                                                                                                                               & Set times, s                                                                                                                                \\ \hline
Matera         & 16.7046        & 40.6486       & 536.9     & \begin{tabular}[c]{@{}c@{}}18798;24489;30561;\\ 68463;74397\end{tabular}                                                                    & \begin{tabular}[c]{@{}c@{}}19116;25230;31131;\\ 69102;75102\end{tabular}                                                                    \\ 
Maspalomas     & -15.6338       & 27.7629       & 205.1     & \begin{tabular}[c]{@{}c@{}}30426;36321;74811;\\ 80661\end{tabular}                                                                          & \begin{tabular}[c]{@{}c@{}}31032;37026;75297;\\ 81396\end{tabular}                                                                          \\ 
Svalbard       & 11.8883        & 78.9067       & 474.0     & \begin{tabular}[c]{@{}c@{}}1203;7167;13125;\\ 19098;25170;31167;\\ 37284;43443;49599;\\ 55704;61755;67782;\\ 73791;79785;85764\end{tabular} & \begin{tabular}[c]{@{}c@{}}1962;7920;13881;\\ 19860;25851;31857;\\ 37878;43920;50010;\\ 56157;62322;68541;\\ 74523;80541;86400\end{tabular} \\ \hline \hline
\end{tabular}
\label{Tab_5}
\end{table}

\section{Conclusions}
This paper has shown a new analytical method for the determination of the coverage area associated to a spacecraft navigation antenna or an optical sensor having a conical field of view when the Earth is modeled as an oblate ellipsoid of rotation. In particular, the satellite’s position vector, the direction of the navigation line of sight, and the half-aperture angle of the conical field of view assumed for the propagation of the signal are required as input data.

The major achievement has been the conversion of the set of equations used to derive the most relevant parameters of the coverage problem (i.e., ground-range angle, elevation angle, horizon-boresight angle) when modeling the Earth’s surface as a sphere into a new set where the shape of the Earth’s surface is an oblate ellipsoid of rotation. Another important aspect has been the definition of a proper analytical formulation to derive an algorithm with low computational time.
The results of the new analytical method have been compared to the ones obtained considering the Earth’s surface modeled as a sphere. Such results show a significant refinement in the determination of the coverage parameters. Finally, the new geometry of the problem allows dealing with different types of pointing scenarios such as the geodetic/nadir pointing configuration.

\appendix
\renewcommand\thefigure{A\arabic{figure}} 
\renewcommand{\theequation}{A\arabic{equation}}
\section*{Appendix: Spherical Model}
\setcounter{figure}{0} 
\setcounter{equation}{0}
\label{appendice}

\begin{figure}[ht!]
	\centering
	\includegraphics[scale = 0.65]{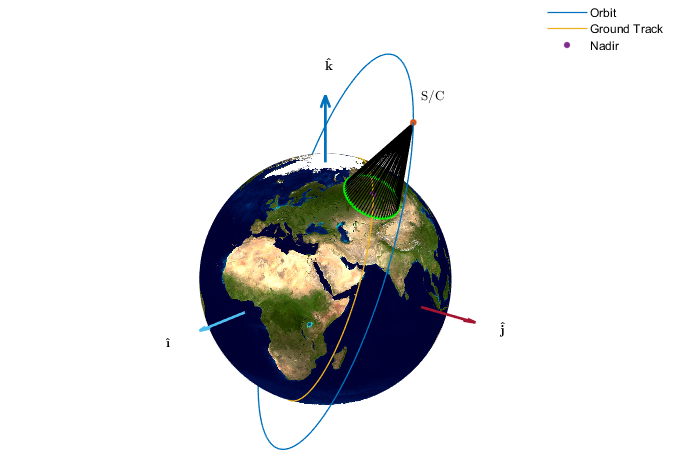}
	\caption{Geometry for the field of view in the Earth-centered Earth-fixed frame.}
	\label{Fig_15}
\end{figure}

The spherical approach is the simplest way to solve the problem of computing the Earth coverage \cite{Vallado}. Indeed, this model assumes the Earth as a perfect sphere, which is a first order approximation. This approximation can be considered good if the level of accuracy and precision required from the problem is in the order of kilometers. Figure \ref{Fig_15} represents the geometry used for the field-of-view (FOV) from a 3D view.

\begin{figure}[ht!]
	\centering
	\includegraphics[scale = 0.9]{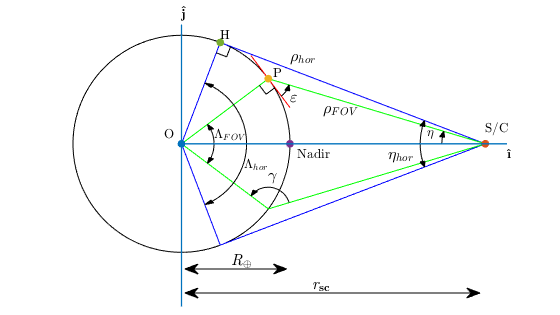}
	\caption{Planar representation of the field-of-view problem.}
	\label{Fig_16}
\end{figure}

The general problem associated with the calculation of the ground range angle ($\Lambda$\textsubscript{FOV}), which is the convex angle measured between the two lines connecting the center of the Earth with the extreme intersections of the signal conical field of view with the Earth's surface (Fig.\hspace{1.5 mm}\ref{Fig_16}), is to determine the amount of surface that can be seen given the satellite's position vector and the half-aperture angle at the satellite. 
If the navigation signal is modeled as a cone with a given half-aperture angle extending from the center of mass of the satellite, the spherical symmetry gives the possibility to study the problem in a planar way as shown in Fig.\hspace{1.5 mm}\ref{Fig_16}.
The point directly below the satellite is called \textit{nadir} point. The half-aperture angle, $\eta$, defines the angular displacement from the nadir direction. From simple geometrical considerations, the maximum coverage area can be determined tracing the two tangents to the sphere from the spacecraft's center of mass.
The horizon-boresight angle, $\eta$\textsubscript{hor}, can be computed considering the right triangle OHS with the following formula:

\begin{equation}
	\sin(\eta\textsubscript{hor}) = \frac{R_{\oplus}}{r_{s/c}}
	\label{Eq_57}
\end{equation}
where $R_{\oplus}$ is the Earth mean equatorial radius and $r$\textsubscript{sat} is the distance of the satellite from the Earth center.
The sine expression is sufficient because this angle will never exceed $\pm90^{\circ}$. Using the same right triangle, also the horizon ground range angle, $\Lambda$\textsubscript{hor}, can be determined using planar trigonometry:

\begin{equation}
\cos(\Lambda\textsubscript{hor}) = \frac{R_{\oplus}}{r_{s/c}}
\label{Eq_58}
\end{equation}

For the slant range to the horizon, $\rho$\textsubscript{hor}, planar trigonometry is also used:

\begin{equation}
\rho\textsubscript{hor} = r_{s/c}\cos(\eta\textsubscript{hor})
\label{Eq_59}
\end{equation}

It is possible to get a general expression for the slant range by examining the oblique triangle SPO in Fig.\hspace{1.5 mm}\ref{Fig_16}. First, the intermediate angle, $\gamma$, is calculated using the sine law for oblique triangles:

\begin{equation}
\sin(\gamma) = \frac{r_{s/c}\sin(\eta)}{R_{\oplus}}
\label{Eq_60}
\end{equation}

In particular, this intermediate angle should be always be higher than $90^{\circ}$.
The generic slant range to any point on the Earth is computed by using the following equation:

\begin{equation}
\rho\textsubscript{FOV} = R_{\oplus}\cos(\gamma) + r_{s/c}\cos(\eta)
\label{Eq_61}
\end{equation}

As a verification, the slant range value is always between the height of the satellite and the slant range to the horizon.
Finally, it is possible to determine the ground range angle using the sine law:

\begin{equation}
\sin\left(\frac{\Lambda\textsubscript{FOV}}{2}\right) = \frac{\rho\textsubscript{FOV}\sin(\eta)}{R_{\oplus}}
\label{Eq_62}
\end{equation}

Also in this case, the ground range angle needs to be less than $\pm90^{\circ}$ and less than the horizon ground range angle.
It is also possible to convert the ground range from an angle to a length by simply multiplying by the Earth radius:

\begin{equation}
\Lambda\textsubscript{km} = R_{\oplus}\Lambda\textsubscript{FOV}
\label{Eq_63}
\end{equation}
where the ground range angle needs to be expressed in radians.
The last quantity of interest is the elevation angle, $\varepsilon$, which is the angle between the tangent to the sphere in a point and the conjunction of the point of tangency with the satellite's center of mass.
Looking at Fig. \ref{Fig_16}, the elevation angle, the ground range angle, and the half-aperture angle are related to each other by the following equation:

\begin{equation}
\varepsilon = \frac{\pi}{2} - \frac{\Lambda\textsubscript{FOV}}{2} - \eta
\label{Eq_64}
\end{equation}

The geographic coordinates of the two points associated to the aperture angle can be obtained and in this way all the important quantities for the coverage analysis are determined.

\section*{Acknowledgment}
The research leading to these results has received funding from the European Research Council (ERC) under the European Union Horizon 2020 research and innnovation program as part of the project COMPASS (Grant agreement No. 679086).

\bibliography{Bibliography}

\end{document}